\documentclass[journal]{IEEEtran}
\usepackage{cite}
\usepackage{amsmath,amssymb,amsfonts}
\usepackage{algorithmicx}
\usepackage{algorithm}
\usepackage{graphicx}
\usepackage{textcomp}
\usepackage{xcolor}
\usepackage{amsmath}
\usepackage{bm}
\usepackage{float}
\usepackage{subcaption}
\usepackage{tikz}
\usepackage{pgflibraryarrows}
\usepackage{pgflibrarysnakes}
\usepackage{xcolor}
\usepackage{tikz-3dplot}
\usepackage[export]{adjustbox}
\usepackage{pgfplots}
\definecolor{meta_col}{rgb}{0, 0, 0}
\definecolor{irs_col}{rgb}{0, 1, 0}
\definecolor{v_col}{rgb}{0,0.4470,0.7410} %[0.4940 0.1840 0.5560]

\def\BibTeX{{\rm B\kern-.05em{\sc i\kern-.025em b}\kern-.08em
		T\kern-.1667em\lower.7ex\hbox{E}\kern-.125emX}}
\makeatletter
\tikzset{
	dot diameter/.store in=\dot@diameter,
	dot diameter=1pt,
	dot spacing/.store in=\dot@spacing,
	dot spacing=0.5pt,
	dots/.style={
		line width=\dot@diameter,
		line cap=round,
		dash pattern=on 0pt off \dot@spacing
	}
}

\begin{document}
\title{Binary Polarization Shift Keying with Reconfigurable Intelligent Surfaces} %%%-.-<-._-_.->---..---<-.-.->-.-%%%
\author{Emad~Ibrahim,
	Rickard~Nilsson, and~Jaap~van~de~Beek% <-this % stops a space
	\thanks{The authors are with the Department of Computer Science, Electrical
		and Space Engineering, Luleå University of Technology, 97187 Luleå,
		Sweden e-mail:\{emad.farouk.ibrahim, rickard.o.nilsson, jaap.vandebeek\}@ltu.se}%%-.....-<-this % stops a space
	}
\maketitle
\begin{abstract}
We propose a novel binary polarization shift keying modulation scheme for a line-of-sight environment by exploiting the polarization control ability of the reconfigurable intelligent surface (RIS). The RIS encodes the information data in terms of the polarization states of either the reflected wave from the RIS or the composite wireless channel between an RF source and receiver. In the first case, polarization mismatch correction becomes essential at the receiver. In the second case, the RIS pre-codes the reflected wave to compensate for the polarization mismatch which allows non-coherent demodulation at the receiver.
\end{abstract}
\section{Introduction}
\let\thefootnote\relax\footnote{We acknowledge the partial funding by the EU's Interreg Nord Program.}
Current research on reconfigurable intelligent surface (RIS) aided wireless communications is mainly focused on the beam focusing and steering ability of the RIS for different communication systems \cite{b1}. However, one of the potential functions of the RIS is that it can also control the polarization state of the reflected waves\cite{b2}. The polarization state of the electromagnetic waves describes the orientation of the electric field relative to the direction of propagation which can be either linear, circular or elliptical. Typically, dual-polarized (DP) antennas of two orthogonal polarized states have been used to provide diversity and multiplexing gains \cite{b3}. Recently, it was shown that RIS polarization manipulation becomes possible thanks to the deployment of DP reflecting elements which could excite two orthogonal polarization states and induce independent phase shifts per each polarization state whenever a wave is incident on them \cite{b4,b5}. 

A promising application for the RIS is to serve as an access point for information transfer. In this case, the RIS exploits an ambient or dedicated radio-frequency (RF) source for information encoding, e.g., the data sources may be some sensors that collect specific measurements and have direct connections to the RIS. In this letter, we propose a novel RIS assisted information transfer scheme by encoding the information data in terms of the polarization state of the reflected waves from the RIS. We assume a line-of-sight (LoS) environment which often exists in many scenarios such as millimetre-wave and rural area communications.

Few works in the open literature have suggested using the RIS for information transfer. In \cite{b6} the authors propose a RIS that alternates the states of its reflecting elements, by turning them on and off, to perform amplitude shift keying (ASK) modulation for information encoding. That, however, adds complexity to the RIS fabrication since the amplitude reflection coefficients of its elements need to be controlled. Also, it underutilizes the RIS since some elements are switched off which decreases the received signal-to-ratio (SNR)\cite{b7}. In \cite{b8} the RIS is used to perform space shift keying modulation, where the RIS encodes the information data in the indices of the receiver antennas by maximizing the received SNR for the selected antenna index. However, a fundamental limit in the space shift keying is that it requires a rich-scattering environment wherein the receiver antennas have low correlation, while LoS environments with highly correlated antennas limit its performance \cite{b9}.

Moreover, few works have discussed the deployment of DP-RIS. In \cite{b10} the authors built a prototype of a DP-RIS to provide a multiplexing gain by encoding two independent streams over the two degrees of freedom of the DP-RIS. In \cite{b11} the authors fabricated a DP-RIS to manipulate the polarization state of the reflected waves and constructed a prototype for a RIS assisted polarization modulation scheme. However, the work in \cite{b11} introduces a fabrication technique for the DP elements in the RIS. Also, it doesn't include the polarization mismatch loss that usually occurs in a LoS environment.

The main contributions are that we first present a mathematical framework for the RIS assisted polarization control functionality. Then, we develop two novel RIS assisted information transfer schemes that exploit the polarization state of the reflected wave from the RIS for information encoding. In one scheme, the RIS is deployed to beam steer the incident wave from a single polarized RF source towards a DP receiver and switch the polarization state of the reflected wave between vertical and horizontal states to encode the information bits. In another scheme, to avoid the polarization mismatch estimation and correction at the receiver, the RIS also pre-codes the reflected wave to compensate for the polarization mismatch which allows non-coherent demodulation such that a simple maximum power detector at the receiver becomes efficient.     
\section{System Model}
\begin{figure}[t!]
	\centering
	\begin{center}
		\tdplotsetmaincoords{90}{90}
		\begin{tikzpicture}[tdplot_main_coords,scale=0.85]
		
		\tdplotsetrotatedcoords{-2}{0}{0}
		\coordinate (Shift) at (0,-3.4,0.2);
		\tdplotsetrotatedcoordsorigin{(Shift)}
		\filldraw[tdplot_rotated_coords,rounded corners,draw=black,fill=blue,line width=0.5pt,opacity=0.65] (0,-0.2,0.1725)--(0,-0.2,0.725)--(0,1.5,0.725)--(0,1.5,0.1725)--cycle;
		\draw[tdplot_rotated_coords,line width=1pt,opacity=0,<->,-latex](0,0.65,0.82)--(0,0.65,0.82) node[anchor=north ,rotate=0,opacity=1,white] {\scriptsize Information};;
		\draw[tdplot_rotated_coords,line width=1pt,opacity=0,<->,-latex](0,0.65,0.53)--(0,0.65,0.53) node[anchor=north ,rotate=0,opacity=1,white] {\scriptsize source};;
		\draw[tdplot_rotated_coords,line width=0.25pt,opacity=1,-stealth](0,1.5,0.45)--(0,2.4,0.45);
		\draw[tdplot_rotated_coords,line width=1pt,opacity=0,<->,-latex](0,1.95,0.37)--(0,1.95,0.37) node[anchor=south ,rotate=0,opacity=1] {\tiny $ 010... $};;
		
		\def \phadiff {2.617}%120deg
		\def \inipoi {\phadiff*1.337/6.283}
		\def \secpoi {0.667}
		%reflecting element
		%
		%\filldraw[color=white,fill=white,opacity=1] (-2.3,0,0) circle (2cm) node[anchor=north west] {$$};

		\def \irsside {1};
		\def \n {6};
		\def \spacing {0.1};
		\def \sidelength {0.2167};
		
		%\tdplotsetrotatedcoords{0}{0}{0}
		\tdplotsetrotatedcoords{0}{0}{0}
		\coordinate (Shift) at (-0,0,0);
		\tdplotsetrotatedcoordsorigin{(Shift)}
		\filldraw[tdplot_rotated_coords,draw=black,fill=lime,line width=1pt,opacity=1] (0,-\irsside,-\irsside)--(0,-\irsside,\irsside)--(0,\irsside,\irsside)--(0,\irsside,-\irsside)--cycle;
		\coordinate (Shift) at (0,-\irsside,-\irsside);
		\tdplotsetrotatedcoordsorigin{(Shift)}
		\foreach \q in {1,...,\n}
		{
			\foreach \s in {1,...,\n}
			{

				\filldraw[tdplot_rotated_coords,fill=yellow,line width=0pt,opacity=1] (0,\s*\spacing+\s*\sidelength-\sidelength,\q*\spacing+\q*\sidelength-\sidelength)--(0,\s*\spacing+\s*\sidelength,\q*\spacing+\q*\sidelength-\sidelength)--(0,\s*\spacing+\s*\sidelength,\q*\spacing+\q*\sidelength)--(0,\s*\spacing+\s*\sidelength-\sidelength,\q*\spacing+\q*\sidelength)--cycle;

				\draw[tdplot_rotated_coords,line width=0.5pt,opacity=1,red](0,\s*\spacing+\s*\sidelength-\sidelength+0.01,\q*\spacing+\q*\sidelength-\sidelength+0.01)--(0,\s*\spacing+\s*\sidelength-0.01,\q*\spacing+\q*\sidelength-0.01);
				\draw[tdplot_rotated_coords,line width=0.5pt,opacity=1,blue](0,\s*\spacing+\s*\sidelength-0.01,\q*\spacing+\q*\sidelength-\sidelength+0.01)--(0,\s*\spacing+\s*\sidelength-\sidelength+0.01,\q*\spacing+\q*\sidelength-0.01);

			}
			
		}
		\coordinate (Shift) at (-0,0,0);
		\tdplotsetrotatedcoordsorigin{(Shift)}
		
		\draw[tdplot_rotated_coords,line width=0.5pt] (0,-0.16,-0.79)  [x={(0,0,1)}] circle (0.17);
		\draw[tdplot_rotated_coords,line width=0.5pt,opacity=1](0,-0.16,-0.95)--(0,-0.16,-1.2) ;
		\draw[tdplot_rotated_coords,line width=0.5pt,opacity=1,-stealth](0,-0.16,-1.2)--(0,0.1,-1.2) ;
		\draw[tdplot_rotated_coords,line width=1pt,opacity=0,<->,-latex](0,+1,-1.45)--(0,1,-1.45) node[anchor=south ,rotate=0,opacity=1] {\tiny Reflecting element};;
		%incident wave
		%\tdplotsetrotatedcoords{0}{0}{0}--tdplot_rotated_coords
		%\draw[line width=0.25pt,opacity=1](tx)--(ris);
		%\filldraw[fill=black,fill=blue!80,opacity=1] (tx) circle (0.05cm) node[anchor=north west] {$Tx$};	
		\tdplotsetrotatedcoords{0}{25}{-35}
		%\coordinate (Shift) at  (-1,-1,1);
		\tdplotsetrotatedcoordsorigin{(Shift)}
		\draw[tdplot_rotated_coords,line width=0.5pt,black,variable=\x,samples at={1,1.01,...,4.49}] plot (\x,0,{-0.707*sin(9*(\x -1) r)});
		
		\foreach \x in{1,1.01,...,4.49}{
			\draw[tdplot_rotated_coords,color=black,line width=0.01 pt, opacity=0.1](\x,0,0) -- (\x,0,{-0.707*sin(9*(\x -1) r)});}
		
		\draw[tdplot_rotated_coords,draw=black,fill=yellow,line width=0.1pt,fill opacity=1] (0,0,0)--(4.49,0,0);
		%,->,-latex
		\draw[tdplot_rotated_coords,line width=0.25pt,opacity=1,-stealth](4.75,0,1)--(2.25,0,1);
		\draw[tdplot_rotated_coords,line width=0pt,opacity=0,->,-latex](5,0,1.6)--(2.4,0,1.6) node[anchor=north east,rotate=30.5,opacity=1] {\scriptsize Incident wave};;
		%%% rf source
		%\tdplotsetrotatedcoords{0}{35}{-33}
		%\coordinate (Shift) at (-1,-1,1);
		\tdplotsetrotatedcoordsorigin{(Shift)}
		\def \txx {5.75};
		
		\filldraw[tdplot_rotated_coords,rounded corners,draw=black,fill=black,line width=0.5pt,opacity=0.5] (\txx,0,0)--(\txx,0,0.5)--(\txx,0.9,0.5)--(\txx,0.9,0)--cycle;
		%%%%vert
		\draw[tdplot_rotated_coords,line width=0.5pt] (\txx,0.45,0.5) -- (\txx,0.45,0.65) ;
		\draw[tdplot_rotated_coords,fill=black,line width=0.5pt] (\txx,0.45,0.65)-- (\txx,0.375,0.8)--(\txx,0.525,0.8)--cycle;
		
		\draw[tdplot_rotated_coords,line width=0pt,opacity=0,->,-latex](\txx,0.45,0.35)--(\txx,0.45,0.35) 
		node[xslant=0.0, yslant=-0.3,rotate=0,opacity=1,color=white]{\scriptsize RF };;
		\draw[tdplot_rotated_coords,line width=0pt,opacity=0,->,-latex,color=white](\txx,0.46,0.1)--(\txx,0.46,0.1) node[xslant=0.0, yslant=-0.3,rotate=0,opacity=1,color=white] {\scriptsize src.};;
		
		% refelected wave
		\tdplotsetrotatedcoords{0}{25}{65} %{0}{35}{-45}
		%\coordinate (Shift) at (-1,-1,1);
		\tdplotsetrotatedcoordsorigin{(Shift)}
		\draw[tdplot_rotated_coords,line width=0.5 pt,v_col,variable=\x,samples at={1,1.001,...,2.396}] plot (\x,{0.707*sin(9*(\x -1) r},0);
		\foreach \x in{1,1.01,...,2.396}{
			\draw[tdplot_rotated_coords,color=v_col,line width=0.01 pt, opacity=0.1] (\x,0,0) -- (\x,{0.707*sin(9*(\x -1) r},0);}
		%
		
		%2
		\draw[tdplot_rotated_coords,line width=0.5 pt,black,variable=\x,samples at={2.396,2.397,...,3.79}] plot (\x,0,{0.707*sin(9*(\x -1) r});
		\foreach \x in{2.396,2.4010,...,3.79}{
			\draw[tdplot_rotated_coords,color=black,line width=0.01 pt, opacity=0.1] (\x,0,0) -- (\x,0,{0.707*sin(9*(\x -1) r});}
		%3
		
		\draw[tdplot_rotated_coords,line width=0.5 pt,v_col,variable=\x,samples at={3.79,3.791,...,5.19}] plot (\x,{0.707*sin(9*(\x -1) r},0);
		\foreach \x in{3.79,3.8,...,5.19}{
			\draw[tdplot_rotated_coords,color=v_col,line width=0.01 pt, opacity=0.1] (\x,0,0) -- (\x,{0.707*sin(9*(\x -1) r},0);}	
		
		\draw[tdplot_rotated_coords,draw=black,fill=yellow,line width=0.2pt,fill opacity=1] (0,0,0)--(5.19,0,0);
		
		%\draw[tdplot_rotated_coords,line width=0.1pt,opacity=1](2.396,0,1.1)--(2.396,0,1.3);%here
		%\draw[tdplot_rotated_coords,line width=0.1pt,opacity=1](3.79,0,1.1)--(3.79,0,1.3);%here

		%\draw[tdplot_rotated_coords,line width=1pt,opacity=0,<->,-latex](1,0,0.9)--(2,0,0.9) node[anchor=south ,rotate=-10,opacity=1] {\tiny $ 0 $};;
		
		%\draw[tdplot_rotated_coords,line width=1pt,opacity=0,<->,-latex](1,0,0.9)--(3.1,0,0.9) node[anchor=south ,rotate=-10,opacity=1,green] {\tiny $ 1 $};;
		
		%\draw[tdplot_rotated_coords,line width=1pt,opacity=0,<->,-latex](3,0,0.9)--(4.5,0,0.9) node[anchor=south ,rotate=-10,opacity=1] {\tiny $ 0 $};;
		
		\draw[tdplot_rotated_coords,line width=0.25pt,opacity=1,-stealth](1.75,0,0.9)--(4.5,0,0.9);
		\draw[tdplot_rotated_coords,line width=0pt,opacity=0,<->,-latex](1.85,0,0.85)--(3.15,0,0.85) node[anchor=south ,rotate=-10,opacity=1] {\scriptsize Reflected wave};;
		
		%%%%receiver
		\tdplotsetrotatedcoordsorigin{(Shift)}
		\def \rxx {5.4};
		\filldraw[tdplot_rotated_coords,rounded corners,draw=black,fill=blue,line width=0.5pt,opacity=0.65] (\rxx,0,-0.6)--(\rxx,0,0)--(\rxx+0.8,0,0)--(\rxx+0.8,0,-0.6)--cycle;

		%%%%vert
		\draw[tdplot_rotated_coords,line width=0.5pt] (\rxx+0.4,0,0) -- (\rxx+0.4,0,0.4) ;
		\draw[tdplot_rotated_coords,fill=black,line width=0.5pt] (\rxx+0.4,0,0.4)-- (\rxx+0.4,-0.1,0.6)--(\rxx+0.4,0.1,0.6)--cycle;
		%%%horz
		\draw[tdplot_rotated_coords,line width=0.5pt] (\rxx+0.4,0,0.2) -- (\rxx+0.4,0.2,0.2) ;
		\draw[tdplot_rotated_coords,fill=black,line width=0.5pt] (\rxx+0.4,0.2,0.2)-- (\rxx+0.4,0.3,0.3)--(\rxx+0.4,0.3,0.1)--cycle;
		
		\draw[tdplot_rotated_coords,line width=0pt,opacity=0,->,-latex](\rxx+0.4,0,-0.175)--(\rxx+0.4,0,-0.175) 
		node[xslant=0.0, yslant=-0.2,rotate=0,opacity=1,color=white]{\scriptsize DP };;
		\draw[tdplot_rotated_coords,line width=0pt,opacity=0,->,-latex,color=black](\rxx+0.395,0,-0.45)--(\rxx+0.395,0,-0.45) node[xslant=0.0, yslant=-0.2,rotate=0,opacity=1,color=white] {\scriptsize Rx};;

		\def \irsside {0.5}
		\def \phadiff {2.617}%120deg
		\def \inipoi {\phadiff*1.337/6.283}
		\def \secpoi {0.667}
		%reflecting element
		\tdplotsetrotatedcoords{0}{0}{0}
		\coordinate (Shift) at (0,1,-1.85);
		%\tdplotsetrotatedcoordsorigin{(Shift)}
		\draw[tdplot_rotated_coords,draw=black,fill=yellow,line width=0pt,fill opacity=1] (0,\irsside,\irsside)--(0,\irsside,-\irsside)--(0,-\irsside,-\irsside)--(0,-\irsside,\irsside)--cycle;
		\draw[tdplot_rotated_coords,line width=0.5pt,opacity=1,red](0,-\irsside,-\irsside)--(0,\irsside,\irsside);
		\draw[tdplot_rotated_coords,line width=0.5pt,opacity=1,blue](0,\irsside,-\irsside)--(0,-\irsside,\irsside);

		%incident wave	
		\tdplotsetrotatedcoords{0}{25}{-35}
		%\coordinate (Shift) at  (-1,-1,1);
		\tdplotsetrotatedcoordsorigin{(Shift)}
		\draw[tdplot_rotated_coords,line width=0.5pt,black,variable=\x,samples at={0,0.01,...,2.359}] plot (\x,0,{0.5*cos(6*(\x ) r)});
		
		\foreach \x in{0,0.01,...,2.359}{
			\draw[tdplot_rotated_coords,color=black,line width=0.01 pt, opacity=0.1] (\x,0,0) -- (\x,0,{0.5*cos(6*(\x ) r)});}
		
		%\foreach \x in{1,1.01,...,6.59}{
		%	\draw[tdplot_rotated_coords,color=black,line width=0.01 pt, opacity=0.2](\x,0,0) -- (\x,0,{-1*sin(4.5*(\x -1) r)});}
		
		\draw[tdplot_rotated_coords,draw=black,fill=yellow,line width=0.1pt,fill opacity=1] (0,0,0)--(2.356,0,0);

		\def \phadiff {2.617}%120deg
		%1st-reflected wave
		\tdplotsetrotatedcoords{0}{25}{65} %{0}{35}{-45}
		\draw[tdplot_rotated_coords,line width=0.5pt,domain=0:2.356,samples=200,red] plot(\x,{0.75*cos((-\phadiff+6*\x) r)},{0.07*cos((-\phadiff+6*\x) r)});
		
		\foreach \x in{0,0.01,...,2.356}{
			\draw[tdplot_rotated_coords,color=red,line width=0.01 pt, opacity=0.1] (\x,0,0) -- (\x,{0.75*cos((-\phadiff+6*\x) r)},{0.07*cos((-\phadiff+6*\x) r)});}

		\draw[tdplot_rotated_coords,line width=0.1pt,opacity=1](0,0,0)--(2.795,0,0) ;
		\draw[tdplot_rotated_coords,line width=0.5pt,domain=0:2.356,samples=200,blue] plot(\x,-{0.65*cos((6*\x) r)},{0.6*cos((6*\x) r)});
		
		\foreach \x in{0,0.01,...,2.356}{
			\draw[tdplot_rotated_coords,color=blue,line width=0.01 pt, opacity=0.1] (\x,0,0) -- (\x,-{0.65*cos((6*\x) r)},{0.6*cos((6*\x) r)});}
		
		\draw[tdplot_rotated_coords,line width=0.5pt,domain=2.35:2.795,samples=50,red,dot spacing=0.5mm,dots] plot(\x,{0.75*cos((-\phadiff+6*\x) r)},{0.07*cos((-\phadiff+6*\x) r)});
		
		\draw[tdplot_rotated_coords,line width=0.2pt,opacity=1](2.33,-0.3,0)--(2.33,0.4,0) ;
		\draw[tdplot_rotated_coords,line width=0.2pt,opacity=1](2.795,-0.3,0)--(2.795,0.4,0) ;
		
		\draw[tdplot_rotated_coords,line width=0.2pt,opacity=1,stealth-stealth](2.33,-0.25,0)--(2.795,-0.25,0);
		%\draw[tdplot_rotated_coords,line width=0.2pt,opacity=1,->,-latex](4.91,0,-0.07)--(4.24,0,-0.07);	
		\draw[tdplot_rotated_coords,line width=0pt,opacity=0,->,-latex](2.7,-0.6,0)--(2.71,-0.6,0) 
		node[xslant=-0.0, yslant=-0.0,rotate=-10,opacity=1,color=black]{\tiny $ \Delta\varphi_{_{m}} $ };;
		
		%\tdplotsetrotatedcoords{0}{0}{0}
		%\draw[tdplot_rotated_coords,line width=0.25pt] (0,0.55,-0.35)  [x={(0,0,1)}] circle (1.05 and 2.25);
		%\draw[tdplot_rotated_coords,line width=0.25pt] (0,0.55,-0.35)  [x={(0,0,1)}] circle (1.07 and 2.27);

		\end{tikzpicture}
	\end{center}
	\vspace*{-3mm}  
	\caption{RIS-aided polarization modulation.} 
	\label{fig:1}
\end{figure}
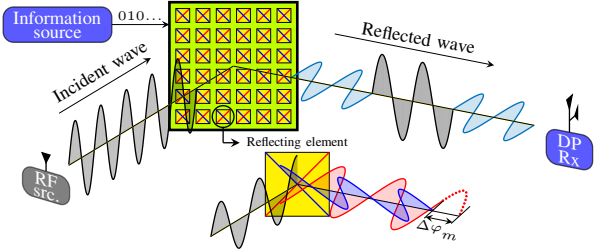

We consider an RIS of $ M $ lossless DP reflecting elements each of which has two polarization states of slant $45^{\circ}$ and slant $-45^{\circ}$ as shown in Fig. \ref{fig:1}. In addition, each element induces two independent phase shifts denoted by $ \varphi_{_{m,1}} $ and $ \varphi_{_{m,2}} $ for the excited slant $45^{\circ}$ and slant $-45^{\circ}$ polarization states, respectively, while $ \varphi_{_{m,1}}$ and $ \varphi_{_{m,2}} \in[0,2\pi] $,  $\forall\,m\in\mathcal{M} $ such that $\mathcal{M}=\left\lbrace1,2,...,M\right\rbrace$. 

\setlength{\textfloatsep}{5pt} 

In this letter, we assume an RF source that transmits a vertically polarized unmodulated carrier wave which is received by a DP receiver of vertically and horizontally polarized antennas solely through the wave reflection on the RIS (the direct channel between the RF source and the receiver is blocked) as shown in Fig. \ref{fig:1}. Thus, the composite wireless channel $\mathbf{h} \in \mathbb{C}^{2{\times}1}$ between the receiver and source becomes 
\begin{equation}%1
\label{1}
\mathbf{h}=\sum_{m=1}^{M}  \mathbf{H}_{2_{m}}\Phi_{m}\mathbf{h}_{1_{m}},
\end{equation}
where $\mathbf{h}=[h_{_{\textnormal{V}}},h_{_{\textnormal{H}}}]^{\textnormal{T}}$ such that $ h_{_{\textnormal{V}}} $ and $ h_{_{\textnormal{H}}}$ are the composite channels from the source to the vertical and horizontal antennas, respectively, while $\mathbf{h}_{1_{m}} \in\mathbb{C}^{2{\times}1} $  and  $ \mathbf{H}_{2_{m}}\in\mathbb{C}^{2{\times}2} $  are the channels between the $ m $th reflecting element in the RIS to source and receiver, respectively. The matrix $\Phi_{m} \in \mathbb{C}^{2{\times}2}$ is diagonal and accounts for the two phase shifts induced by the $ m $th reflecting element as
\begin{equation}%2
\label{2}
\Phi_{m}=\begin{bmatrix}  e^{j\varphi_{_{m,1}}} & 0\\0 & e^{j\varphi_{_{m,2}}} \end{bmatrix}.
\end{equation}

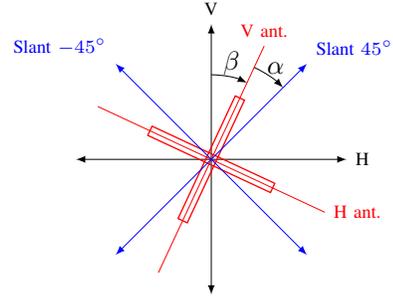
\begin{figure}[t!]
	\centering
	\begin{center}
		\tdplotsetmaincoords{0}{0}
		\begin{tikzpicture}[tdplot_main_coords,scale=0.9]
		\draw[line width=0.25pt,opacity=1,<->,latex-latex](-2,0)--(2,0) node[anchor=west] {\scriptsize H};
		\draw[line width=0.25pt,opacity=1,<->,latex-latex](0,-2)--(0,2) node[anchor=south] {\scriptsize V};
		
		\tdplotsetrotatedcoords{-25}{0}{0}
		\draw[tdplot_rotated_coords,line width=0.5pt,opacity=1,color=red](-0.075,-1)--(-0.075,1)--(0.075,1)--(0.075,-1)--cycle;
		\draw[tdplot_rotated_coords,line width=0.25pt,opacity=1,color=red](0,-1.85)--(0,1.85) node[anchor=south] {\scriptsize V ant.};
		
		\tdplotsetrotatedcoords{65}{0}{0}
		\draw[tdplot_rotated_coords,line width=0.5pt,opacity=1,color=red](-0.075,-1)--(-0.075,1)--(0.075,1)--(0.075,-1)--cycle;
		\draw[tdplot_rotated_coords,line width=0.25pt,opacity=1,color=red] (0,1.85)--(0,-1.85)  node[anchor=west] {\scriptsize H ant.};
		\tdplotdrawarc[line width=0.25pt,->,>=latex]{(0,0,0)}{1.25}{90}{65}{anchor= east}{$$};
		\node (A) at (0.3,1.43,0) {$\beta$};
		\tdplotdrawarc[line width=0.25pt,->,>=latex]{(0,0,0)}{1.5}{65}{45}{anchor= east}{$$};
		\node (A) at (0.95,1.35,0) {$\alpha$};
		\tdplotsetrotatedcoords{45}{0}{0}
		
		\draw[tdplot_rotated_coords,line width=0.25pt,opacity=1,<->,latex-latex,color=blue](-2,0)--(2,0) node[anchor= south west] {\scriptsize  Slant $45^{\circ}$};	
		\draw[tdplot_rotated_coords,line width=0.25pt,opacity=1,<->,latex-latex,,color=blue](0,-2)--(0,2) node[anchor=south east] {\scriptsize  Slant $-45^{\circ}$};
		\end{tikzpicture}
	\end{center}
	\vspace*{-3mm}  
	\caption[short text]{Polarization mismatch between the reflected wave and the DP receiver.}
	\label{fig:2}
\end{figure}

In a LoS environment, orthogonality between two polarized states is maintained through the wireless channel. However, polarization mismatch loss often occurs due to the different orientations between the DP received wave and destination\cite{b12}. In this paper, we assume that the first-hop channel between the RF source and the RIS is perfectly aligned such that there is no polarization mismatch loss. However, there is a polarization mismatch loss in the second-hop channel between the RIS and the receiver. In addition, we assume that the polarization mismatch loss occurs solely due to an azimuth angle mismatch between reflected wave by the RIS and the DP receiver as shown in Fig. \ref{fig:2}. Consequently, the second-hop channel between the $m$th reflecting element of slant $45^{\circ}/-45^{\circ}$ polarization states and the DP receiver of vertically and horizontally polarized antennas becomes \cite{b12}
\begin{equation}%3
\label{3}
\mathbf{H}_{2_{m}}=\rho_{_{2}} e^{j\mu_{_2,m}} \begin{bmatrix} \cos\left(\alpha\right) & \sin\left(\alpha\right)\\\sin\left(\alpha\right) & -\cos\left(\alpha\right) \end{bmatrix},
\end{equation}
where $ \rho_{_{2}} $ and $ \mu_{_2,m} $ account for the large-scale fading channel and the phase shift of the link between the $m$th reflecting element and the receiver, respectively, whereas $\alpha=45^{\circ}-\beta$ is the mismatch angle between the slant $45^{\circ}$ reflected wave and the vertical antenna of the receiver such that $ \beta $  represents the polarization mismatch angle of the DP receiver as shown in Fig. \ref{fig:2}. Similarly, the first-hop channel between the vertical polarized RF source and the slant  $45^{\circ}/-45^{\circ}$  $ m $th reflecting element, given there is no polarization mismatch, becomes $ \mathbf{h}_{1_{m}}=\frac{\rho_{_{1}}}{\sqrt{2}} e^{j\mu_{_1,m}}\left[1,1\right] ^{\textnormal{T}} $ where $ \rho_{_{1}} $ and $ \mu_{_1,m} $ account for the large-scale fading channel and the phase shift, respectively. Furthermore, it is important to note that the large-scale fading channels and the polarization mismatch angle are assumed constant for all the elements in the RIS which is valid assumption given that the RIS is placed in the far-field with respect to both the RF source and the receiver. Consequently, the composite channel matrix in \eqref{1} can be simplified as
\begin{equation}%4
\label{4}
\mathbf{h}= \mathbf{A} \mathbf{u},
\end{equation}
where $ \mathbf{A} $ represents the polarization mismatch loss matrix of the DP receiver which is defined as\cite{b12}
\begin{equation}%5
\label{5}
\mathbf{A}=\begin{bmatrix} \cos\left(\beta\right) & \sin\left(\beta\right)\\-\sin\left(\beta\right) & \cos\left(\beta\right) \end{bmatrix},
\end{equation}
and $ \mathbf{u} $ represents the composite reflected wave from all the elements in the RIS, 
\begin{equation}%6
\label{6}
\mathbf{u}=\dfrac{\eta}{2} \sum\limits_{m=1}^{M} e^{j\left( \psi_{_{m}}+\varphi_{_{m,1}}\right)} 
\begin{bmatrix} 
	1+e^{j\Delta\varphi_{_{m}}}\\
	1-e^{j\Delta\varphi_{_{m}}} 
\end{bmatrix},
\end{equation}
where $ \eta=\rho_{_{1}}\rho_{_{2}} $, $ \psi_{_{m}}=\mu_{_{1,m}}+\mu_{_{2,m}} $, and $ \Delta\varphi_{_{m}}=\varphi_{_{m,2}}-\varphi_{_{m,1}}$ represent the effective large-scale fading, the phase shift of the two-hop channel through the $ m $th element, and the phase shift difference between the two excited polarized states from the $ m $th element, respectively as shown in Fig. \ref{fig:1}. Thus, it is clear that the polarization state of the reflected wave from each reflecting element can be controlled by the appropriate choice of $ \Delta\varphi_{_{m}} $. Particularly, in case of $ \Delta\varphi_{_{m}}=0$ and $ \Delta\varphi_{_{m}}=\pi$ the reflected wave from the $ m $th reflecting element will be vertically and horizontally polarized, respectively. In addition, right and left hand circular polarized waves can be composed when $ \Delta\varphi_{_{m}}=\pi/2$ and $ \Delta\varphi_{_{m}}=-\pi/2$, respectively, and even elliptical polarized waves can be composed for other phase shift difference values.

\setlength{\textfloatsep}{5pt}  
 
\section{RIS based binary Polarization Shift Keying}
In this section, we discuss a modulation scheme where the polarization state of the electromagnetic waves is used to carry information data bits. We develop two different schemes that both exploit polarization for information encoding. In a first scheme, the RIS encodes the information data bits in terms of the polarization states of the reflected wave from the RIS without respect to the polarization mismatch loss. Consequently, in this scheme, the polarization mismatch estimation and correction at the receiver becomes essential for the data detection. In a second scheme, the RIS encodes the information data bits in terms of the polarization states of the composite wireless channel between the RF source and the receiver via the RIS. Consequently, in this scheme, there is no need for the polarization mismatch estimation and correction at the receiver since the RIS itself pre-codes the reflected wave to correct for the polarization mismatch. Moreover, to show the upper bound performance of the proposed schemes we assume that the channels from the RIS to the source and receiver in addition to the polarization mismatch angle are perfectly known. In practical scenarios, an estimation process will be necessary which is possible using pilot signals of limited overhead especially in the LoS environment.
\subsection{Scheme 1: Polarization mismatch correction at the receiver}
In the first scheme, the RIS is deployed to perform two fundamental functions. Firstly, the RIS beam-steers the reflected wave towards the receiver to maximize the composite channel power. Secondly, the RIS shifts the polarization state of the reflected wave between vertical and horizontal polarization states to encode the information data bits as shown in \mbox{Fig. \ref{fig:1}}. Since the composite channel power is maximized whenever all the reflected paths from the RIS are added constructively at the receiver, by observing \eqref{6}, the phase shift of the excited slant $45^{\circ}$ polarization state becomes
\begin{equation}%7
\label{7}
\quad\quad\varphi_{_{m,1}}=-\psi_{_{m}}\quad\quad\;\forall\,m\in\mathcal{M}. 
\end{equation}

In this letter we choose the vertically polarized state to encode information data bit $b=1$ and the horizontally polarized state to encode the information data bit $b=0$. Thus, the phase shift of the excited slant $-45^{\circ}$ polarization state as a function of the information data bits becomes
\begin{equation}%8
\label{8}
\varphi_{_{m,2}}=\varphi_{_{m,1}}+\pi\left(1-b\right) \quad\;\forall\,m\in\mathcal{M}, 
\end{equation}
where \eqref{8} guarantees that all the reflecting elements in the RIS excite vertically and horizontally polarized waves in case of $ b=1 $ and $ b=0 $, respectively. Consequently, the received signal denoted by $\mathbf{y} \in \mathbb{C}^{2{\times}1}$ at the DP receiver as a function of the information data bits becomes
\begin{equation}%9
\label{9}
\mathbf{y}=M\eta\sqrt{p_{_{t}}}\mathbf{A}\mathbf{x}+\mathbf{w},
\end{equation}
where $\mathbf{y}=[y_{_{\textnormal{V}}},y_{_{\textnormal{H}}}]^{\textnormal{T}}$ such that $ y_{_{\textnormal{V}}} $ and $ y_{_{\textnormal{H}}}$ are the received signal at the vertically and horizontally polarized antennas, respectively. The vector $ \mathbf{x}=\left[b,1-b\right]^{\textnormal{T}} $ describes the selected polarization state of the reflected wave as a function of the information data bit, $ p_{_{t}} $ is the transmitted power from the RF source, and $ \mathbf{w} \in \mathbb{C}^{2{\times}1} \sim\mathcal{N_{C}}\left(0,\sigma^2 \mathbf{I}_{2}\right)$ is the additive white Gaussian noise with variance $ \sigma^{2} $. Given the orthogonal polarization mismatch loss matrix shown in \eqref{5}, the equalization process to correct the polarization mismatch simply becomes 
\begin{equation}%10
\label{10}
\mathbf{\hat{x}}=\mathbf{A}^{\textnormal{T}}\mathbf{y}.
\end{equation} 

Then, the data bit demodulation can be performed based on the maximum power detector as
\begin{equation}%11
\label{11}
\hat{b}=
\begin{cases}
1 & \text{ if }\left|\hat{x}_{_{\textnormal{V}}}\right|\geq \left|\hat{x}_{_{\textnormal{H}}}\right|\\
0 & \text{ if }\left|\hat{x}_{_{\textnormal{V}}}\right|<\left|\hat{x}_{_{\textnormal{H}}}\right|
\end{cases}.
\end{equation}

The received SNR at the targeted polarized antenna becomes
\begin{equation}%12
\label{12}
\gamma_{_{1}}=\frac{M^{2}\eta^{2}p_{_{t}}}{\sigma^{2}},
\end{equation}
and the theoretical performance of this scheme in terms of the bit-error-rate (BER) becomes \cite{b13}
\begin{equation}%13
\label{13}
\textnormal{BER}=\frac{1}{2}e^{-\frac{\gamma_{_{1}}}{2}}.
\end{equation}

\subsection{Scheme 2: Polarization mismatch correction at the RIS}
\label{Scheme 2}
In this scheme, the information data bit are encoded in terms of the polarization states of the composite wireless channel between the RF source and receiver via the RIS. Therefore, the RIS performs an additional function in comparison to the previous scheme which is the  pre-coding of the reflected wave to correct the polarization mismatch. 

Let $ l_{_{\textnormal{V}}} $ and $ l_{_{\textnormal{H}}} $ reflecting elements excite solely vertically and horizontally polarized wave, respectively, such that $ \Delta\varphi_{_{m}}=~0,\; \forall\,m\in\mathcal{M}_{_{\textnormal{V}}} $ and $ \Delta\varphi_{_{m}}=\pi,\; \forall\,m\in\mathcal{M}_{_{\textnormal{H}}} $, where $\mathcal{M}_{_{\textnormal{V}}}=\left\lbrace1,2,...,l_{_{\textnormal{V}}}\right\rbrace$ and $\mathcal{M}_{_{\textnormal{H}}}=\left\lbrace l_{_{\textnormal{V}}}+1,l_{_{\textnormal{V}}}+2,...,M\right\rbrace$, while $ l_{_{\textnormal{V}}}+l_{_{\textnormal{H}}}=M $. Consequently, the reflected wave from the RIS defined in \eqref{6} becomes
\begin{equation} %14
\label{14}
\mathbf{u}=\eta
\begin{bmatrix} 
\sum\limits_{m\in\mathcal{M}_{_{\textnormal{V}}}} e^{j(\psi_{m}+\varphi_{_{m,1}})}\\
\sum\limits_{m\in\mathcal{M}_{_{\textnormal{H}}}} e^{j(\psi_{m}+\varphi_{_{m,1}})}
\end{bmatrix}=\eta\begin{bmatrix} 
a_{_{\textnormal{V}}}\\ \\
a_{_{\textnormal{H}}}
\end{bmatrix},
\end{equation}
and the composite channel can be formulated as 
\begin{equation} %15
\label{15}
\mathbf{h}=\eta\cos\left(\beta\right)
\begin{bmatrix} 
a_{_{\textnormal{V}}}+a_{_{\textnormal{H}}}\tan\left(\beta\right) \\
a_{_{\textnormal{H}}}-a_{_{\textnormal{V}}}\tan\left(\beta\right)
\end{bmatrix}. 
\end{equation}

Then, the phase shift of the excited slant $45^{\circ}$ polarization state is chosen to maximize the composite channel power at the targeted polarization state which can be performed based on \eqref{14} and \eqref{15} as 
\begin{equation}%16
\label{16}
\varphi_{_{m,1}}=
\begin{cases}
-\psi_{_{m}}+\dfrac{\pi}{2}[1+(-1)^{b}\textnormal{sign}(\tan(\beta))]\,\forall\,m\in\mathcal{M}_{_{\textnormal{V}}}\\
-\psi_{_{m}}\,\forall\,m\in\mathcal{M}_{_{\textnormal{H}}}  
\end{cases},
\end{equation} 
where sign$ (\cdot) $ denotes the sign operator, while \eqref{16} guarantees the coherent addition for the components at the targeted polarized antenna. Thus, the vertically and horizontally polarized composite channel power, as a function of the information data bit, becomes
\begin{equation}%17
\label{17}
\begin{aligned}
&\qquad\left|h_{_{\textnormal{V}}}\right| ^{2}=\eta^{2}\cos^{2}\left(\beta\right)\left[ l_{_{\textnormal{V}}}-(-1)^{b}l_{_{\textnormal{H}}}\left| \tan\left(\beta\right)\right|\right] ^{2} \\
&\qquad\left| h_{_{\textnormal{H}}}\right| ^{2}=\eta^{2}\cos^{2}\left(\beta\right)\left[l_{_{\textnormal{H}}}+(-1)^{b}l_{_{\textnormal{V}}}\left| \tan\left(\beta\right)\right|\right] ^{2} 
\end{aligned}.
\end{equation}

Moreover, $ l_{_{\textnormal{V}}}$ and $ l_{_{\textnormal{H}}}$ are chosen such that the composite channel vanishes on the horizontally and vertically polarized channels in case of $ b=1 $ and $ b=0 $, respectively, which by observing \eqref{17} becomes
\begin{equation}%18
\label{18}
l_{_{\textnormal{V}}}=
\begin{cases}
\lfloor \frac{M}{1+\left| \tan\left(\beta\right)\right|}\rceil&\text{if $b=1$}\\
\lfloor \frac{M}{1+\left| \cot\left(\beta\right)\right|}\rceil& \text{if $b=0$}
\end{cases},
\end{equation}
where $\lfloor.\rceil$ is the operator that rounds to the nearest integer number, while $ l_{_{\textnormal{H}}}=M-l_{_{\textnormal{V}}} $. Now, given that \eqref{16} and \eqref{18} are fulfilled the data bit detection can be performed directly based on the maximum power detector of the DP received signal as 
\begin{equation}%19
\label{19}
\hat{b}=
\begin{cases}
1 & \text{ if }\left|y_{_{\textnormal{V}}}\right|\geq \left|y_{_{\textnormal{H}}}\right|\\
0 & \text{ if }\left|y_{_{\textnormal{V}}}\right|<\left|y_{_{\textnormal{H}}}\right|
\end{cases}.
\end{equation}

Moreover, by substituting \eqref{18} in \eqref{17} the received SNR at the targeted polarized antenna becomes 
\begin{equation}%20
\label{20}
\gamma_{_{2}}\approx \frac{M^{2}\eta^{2}p_{_{t}}}{\sigma^{2}\left[\;1+\left| \sin\left(2\beta\right)\right|\;\right] }  ,
\end{equation}
where the approximation is due to the rounding operation in \eqref{18}. It is important to note that the received SNR in this scheme degrades in comparison to the previous scheme where the equalization is performed in the receiver. Thus, the RIS correction for the polarization mismatch comes at the expense of the performance. However, this scheme allows for non-coherent data bit detection without the need of polarization mismatch estimation and correction at the receiver which simplifies the receiver structure. The theoretical BER in this scheme becomes\cite{b13}
\begin{equation}%21
\label{21}
\textnormal{BER}\approx\frac{1}{2}e^{-\frac{\gamma_{_{2}}}{2}}.
\end{equation}
\section{Simulation Results}
The simulation parameters are shown in Table \ref{table:1}. We rely on the plate scattering-path loss model which is used given the far-field operation of the RIS as shown in \cite{b14} and we use the radiation pattern of the reflecting element introduced therein. Thus, the large-scale fading channel becomes
\begin{equation}%22
\label{22}
\eta=\left( \dfrac{\Delta\sqrt{G_{t}G_{r}}}{4\pi r_{_{1}}r_{_{2}}}\right) \left[ \cos(\zeta_{_{1}})\cos(\zeta_{_{2}})\right] ^{q_{_{o}}} ,
\end{equation}
where $q_{_{o}}=0.285 $ given the square shape reflecting element of half-wavelength as a side length \cite{b14}, while $ G_{t} $, $ G_{r} $, and $ \Delta $ are the gain of the RF source antenna, the gain of the receiver antennas, and the physical area of the reflecting element, respectively while, $r_{_{1}}$ and $ r_{_{2}}$ represent the distances from the RIS to the source and destination, respectively. Furthermore, $\zeta_{_{1}}$ and $ \zeta_{_{2}}$ represent the angles between the normal of the RIS to the incident and reflected waves, respectively. Importantly, the received SNRs in \eqref{12} and \eqref{20} that govern the performance of the schemes are proportional to $ A_{\textnormal{RIS}}^{2}  $ where  $ A_{\textnormal{RIS}}=M\Delta $ denotes the surface area of the RIS. Thus, in our simulations, we rely on the RIS' surface area instead of its number of elements to neutralize the effect of the carrier frequency on the performance.  

Furthermore, the phase shifts of the links from the reflecting element in the RIS to the source and the receiver are defined based on the plane wave propagation model as      
\begin{equation}%23
\label{23}
\mu_{_{l,m}}=\mathbf{g}_{m}^\mathrm{T}\mathbf{q}_{_{_{l}}} \quad \forall\,m\in\mathcal{M}, 
\end{equation}
where $ l\in\left\lbrace 1,2\right\rbrace $, and
\begin{gather}%24
\label{24}
\mathbf{q}_{_{_{l}}}=\frac{2\pi}{\lambda_{c}}\left[ 
\cos\left(\phi_{_{l}}\right)\cos\left(\theta_{_{l}}\right),\sin\left(\phi_{_{l}}\right)\cos\left(\theta_{_{l}}\right),\sin\left(\theta_{_{l}}\right) \right]^{\textnormal{T}},
\end{gather}
where $ \lambda_{c} $, $\mathbf{g}_{m} \in\mathbb{R}^{3{\times}1} $, and $\mathbf{q}_{_{_{l}}}\in \mathbb{R}^{3{\times}1}$ are the carrier wave-length, the Cartesian coordinates of the $m$th reflecting element in the RIS, and the wave vector which describe the phase variations over the elements in the RIS for the incident/reflected wave, respectively, whereas $\theta_{_{l}}$ and $\phi_{_{l}}$ are the RIS' elevation and azimuth angles of arrival/departure, respectively.

\setlength{\textfloatsep}{5pt}  
\begin{table}[t]
	\label{table}
	\centering
	%\vspace*{-5mm}
	\caption{Simulation Parameters}
	\begin{tabular}{r|r}
		%\hline G_{t}
		\textbf{Parameter}&\textbf{Value}  \\ \hline
		Gain of transmit and receive antennas &$ 3 $ dBi \\ %\hline
		Carrier frequency& $ 3 $ GHz \\ %\hline
		Transmission power ($p_{_{t}}$)&8 dBm \\ %\hline
		Noise power ($\sigma^2$)&-96 dBm \\ %\hline
		Reflecting element aperture area ($\Delta$)&$ \lambda_{c}/2 \times \lambda_{c}/2 $ \\ %\hline
		Location of RF source&$ \left[50, 0, 0\right]m$  \\ %\hline
		Location of Receiver &$ \left[50, 100, 0\right]m$  \\ %\hline
		Location of RIS &$ \left[0, 50, 0\right]m$  
		\label{table:1}
	\end{tabular}	
\end{table}

\setlength{\textfloatsep}{3pt}  
\begin{figure}[b]
	\centering
	\includegraphics[width=8cm,height =4cm]{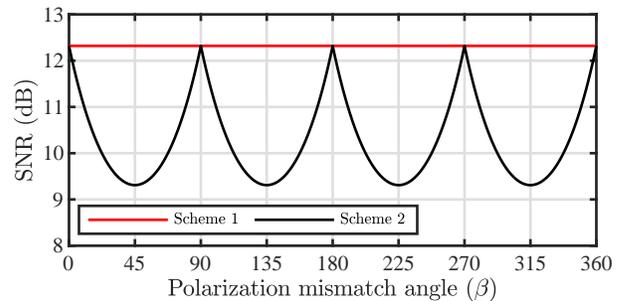}
	\vspace*{-2mm} 
	\caption[short text]{SNR versus the polarization mismatch angle for a RIS of $ A_{\textnormal{RIS}}=1m^{2} $.}
	\label{fig:3}
\end{figure}

\setlength{\textfloatsep}{3pt}  
\begin{figure}[t]
	\centering
	\includegraphics[width=1\columnwidth]{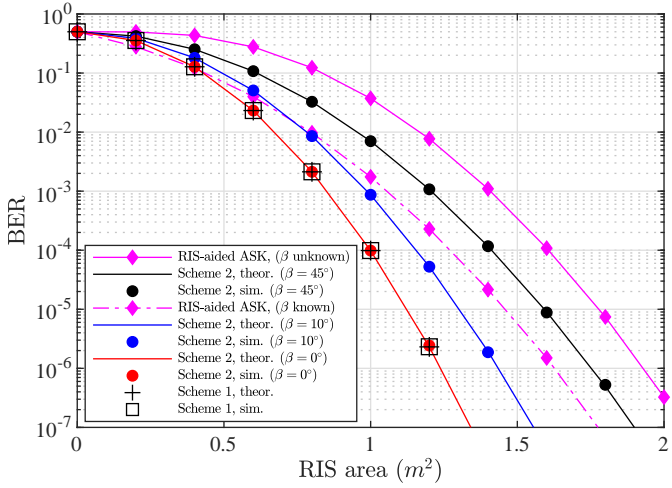}
	\vspace*{-6mm} 
	\caption[short text]{ BER are shown versus the surface area of the RIS.}
	\label{fig:4}
\end{figure}

\setlength{\textfloatsep}{3pt}  
\begin{figure}[b]
	\centering
	\includegraphics[width=1\columnwidth]{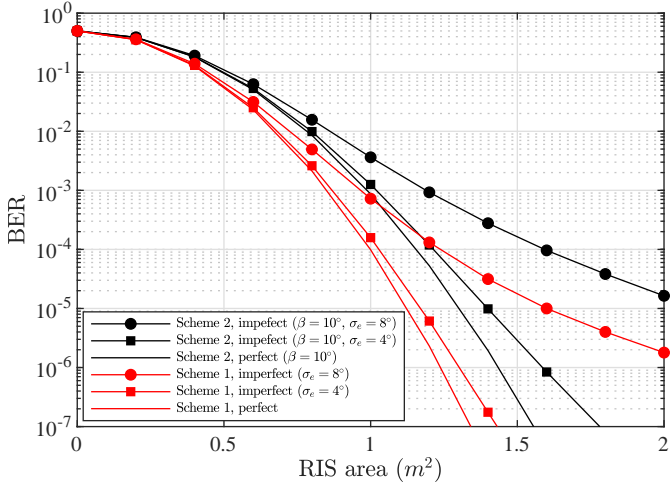}
	\vspace*{-6mm} 
	\caption[short text]{ BER are shown versus the surface area of the RIS in the presence of estimation errors.}
	\label{fig:5}
\end{figure}

In Fig. \ref{fig:3}, the received SNRs at the targeted polarized antenna are shown for the two proposed schemes versus the polarization mismatch angle for a RIS of $ A_{\textnormal{RIS}}=1m^{2} $. For scheme 1, the SNR is independent of the polarization mismatch angle, whereas in scheme 2 the worst performance of $ 3 $ dB loss occurs in case of $ \beta=45^{\circ} $. In addition, the SNR curve is periodic with period equals to $90^{\circ} $, and scheme 2 achieves the same performance as scheme 1 when $\beta$ equals multiples of $90^{\circ}$ and when $\beta=0^{\circ}$. Moreover, it is clear from the performance of scheme 2 that the RIS correction for the polarization mismatch comes at the expense of a reduced performance relative to the polarization mismatch angle. However, scheme 2 allows non-coherent detection which eliminates the need for the polarization mismatch estimation and correction at the receiver in contrast to scheme 1.

In Fig. \ref{fig:4}, the theoretical and simulated BER curves for the two proposed schemes are shown versus the surface area of the RIS. In case of scheme 2 the performance is shown for three different polarization mismatch angles of $ \beta=0^{\circ}, 10^{\circ} $ and $ 45^{\circ} $. It is clear that scheme 2 achieves its best performance when $ \beta=0^{\circ} $ followed by $ \beta=10^{\circ} $ and $ \beta=45^{\circ}$. In addition, scheme 2 achieves similar performance to scheme 1 only when $ \beta=~0^{\circ} $. Furthermore, the performance of the two proposed schemes are compared with the traditional RIS information transfer scheme \cite{b6} where the RIS alternates the on and off states of the reflecting elements to perform ASK modulation. Thus, in case of $ b=0$, all the elements are set to the off states, whereas in the case of $b=1$,  all the elements are set to on states such that the phases of the two excited polarization states are chosen to maximize the received SNR at the DP receiver by fulfilling \eqref{7} and by setting $ \Delta\varphi_{_{m}}=\Delta\varphi \; ,\forall\,m\in\mathcal{M} $  where the performance is independent of the choice of  $ \Delta\varphi \in[0,2\pi] $. In this scenario, the received SNR in case of $ b=1 $ is similar to that in \eqref{12}. Moreover, two possible detectors are performed depending on the knowledge of $ \beta $ at the receiver. The first is the matched filter detector as $ \tilde{x}_{_{ASK,1}}=\left( {\mathbf{h}_{o}^{\textnormal{H}}}/{\left|\mathbf{h}_{o} \right| }\right)   \mathbf{y} $ where $\mathbf{h}_{o}=\mathbf{A}[1+e^{j\Delta\varphi},1-e^{j\Delta\varphi}]^{\textnormal{T}}$, followed by a hard decision:  $\tilde{x}_{_{ASK,1}}$\scriptsize${\underset{\hat{b}=0} {\overset{\hat{b}=1}{\gtrless}}}$\normalsize$M \eta \sqrt{p_{_{t}}}/2$. Alternatively, in case of $ \beta $ is unknown at the receiver a simple possible detector becomes $ \tilde{x}_{_{ASK,2}}=\sqrt{\left|y_{_{\textnormal{V}}} \right|^{2} +\left|y_{_{\textnormal{H}}} \right|^{2}} $ followed by a hard decision similar to the first detector. It is clear from Fig. \ref{fig:4} that through the reasonable performance region, i.e. BER$<10^{-1}$, the proposed \mbox{scheme 1} achieves the best performance, while the performances of the scheme 2 and that of RIS-aided ASK with matched filter detector are comparable depending on the value of $ \beta $. The worst performance is for the RIS-aided ASK when the $ \beta$ is unknown at the receiver.
%\footnotesize$ {\underset{\hat{b}=0} {\overset{\hat{b}=1}{\gtrless}}}$\normalsize $ {\underset{\hat{b}=0} {\overset{\hat{b}=1}{\gtrless}}}$
%$\tilde{x}_{_{ASK,1}} \underset{\hat{b}=0} {\overset{\hat{b}=1}{\gtrless}} M \eta \sqrt{p_{_{t}}}/2$%

In Fig. \ref{fig:5}, the BER performance of the two proposed schemes are presented given there are estimation errors in the polarization mismatch angle. The estimation errors are modeled as a zero-mean Gaussian distribution of standard deviation denoted by $ \sigma_{e} $ which accounts for the accuracy of the estimator. In the case of $ \sigma_{e}=4^{\circ} $, the degradation in the performance are not significant, whereas in the case of  $ \sigma_{e}=8^{\circ} $ the degradation are noticeable. Thus, even imperfect estimator of the polarization mismatch angle can be useful, especially if error coding techniques are used.
\section{Conclusion}
We propose a novel RIS aided binary polarization shift keying modulation method for the LoS environment. Two schemes that exploit the polarization states of either the reflected wave from the RIS or of the composite channel between the RF source and receiver to encode the information data bits are presented. In the first scheme, the receiver corrects the polarization mismatch which occurs in the wireless channel. In the second scheme, the RIS itself pre-codes the reflected wave to compensate for the polarization mismatch. Although this comes at the expense of a degraded performance, it allows non-coherent demodulation without the need for polarization mismatch estimation and correction at the receiver.
%Eventually, the work in this letter opens up a future research direction which is the exploitation of the RIS polarization property to develop higher-order modulation and obtain multiplexing and diversity gains in different propagation environments.


\begin{thebibliography}{00}
\bibitem{b1} M. Di Renzo et al., "Smart Radio Environments Empowered by Reconfigurable Intelligent Surfaces: How It Works, State of Research, and The Road Ahead," in IEEE Journal on Selected Areas in Communications, vol. 38, no. 11, pp. 2450-2525, Nov. 2020, doi: 10.1109/JSAC.2020.3007211.
\bibitem{b2} C. Huang et al., "Holographic MIMO Surfaces for 6G Wireless Networks: Opportunities, Challenges, and Trends," in IEEE Wireless Communications, vol. 27, no. 5, pp. 118-125, October 2020, doi: 10.1109/MWC.001.1900534.
\bibitem{b3} C. Guo, F. Liu, S. Chen, C. Feng and Z. Zeng, "Advances on Exploiting Polarization in Wireless Communications: Channels, Technologies, and Applications," in IEEE Communications Surveys \& Tutorials, vol. 19, no. 1, pp. 125-166, Firstquarter 2017, doi: 10.1109/COMST.2016.2606639.
%DP RIS
\bibitem{b4} Yang, H., Cao, X., Yang, F. et al. A programmable metasurface with dynamic polarization, scattering and focusing control. Sci Rep 6, 35692 (2016). https://doi.org/10.1038/srep35692
\bibitem{b5} S. Sugiura, Y. Kawai, T. Matsui, T. Lee and H. Iizuka, "Joint Beam and Polarization Forming of Intelligent Reflecting Surfaces for Wireless Communications," in IEEE Transactions on Vehicular Technology, vol. 70, no. 2, pp. 1648-1657, Feb. 2021, doi: 10.1109/TVT.2021.3055237.
% RIS assisted ASK
\bibitem{b6} W. Yan, X. Yuan and X. Kuai, "Passive Beamforming and Information Transfer via Large Intelligent Surface," in IEEE Wireless Communications Letters, vol. 9, no. 4, pp. 533-537, April 2020, doi: 10.1109/LWC.2019.2961670.
%Drawback of RIS assisted ASK
\bibitem{b7} Q. Li, M. Wen and M. Di Renzo, "Single-RF MIMO: From Spatial Modulation to Metasurface-Based Modulation," in IEEE Wireless Communications, vol. 28, no. 4, pp. 88-95, August 2021, doi: 10.1109/MWC.021.2000376.
%%%%RIS assisted SSK
\bibitem{b8} E. Basar, "Reconfigurable Intelligent Surface-Based Index Modulation: A New Beyond MIMO Paradigm for 6G," in IEEE Transactions on Communications, vol. 68, no. 5, pp. 3187-3196, May 2020, doi: 10.1109/TCOMM.2020.2971486.
%Draw back of SSK
\bibitem{b9} M. D. Renzo, H. Haas, A. Ghrayeb, S. Sugiura, and L. Hanzo, “Spatial modulation for generalized MIMO: challenges, opportunities, and implementation,” IEEE Proc., vol. 102, no., 1, pp. 56–103, Jan. 2014.
%DP-RIS
\bibitem{b10} X. Chen et al., "Design and Implementation of MIMO Transmission Based on Dual-Polarized Reconfigurable Intelligent Surface," in IEEE Wireless Communications Letters, vol. 10, no. 10, pp. 2155-2159, Oct. 2021, doi: 10.1109/LWC.2021.3095172.
\bibitem{b11} C. X. Huang, J. Zhang, Q. Cheng, and T. J. Cui, “Polarization modulation for wireless communications based on metasurfaces,” Advanced Functional Materials, vol. 31, no. 36, 2103379, Jun. 2021.

%%%Polarization mismatch loss
\bibitem{b12} L. Jian, L. Thiele, and V. Jungnickel, “On the modelling of polarized MIMO channel,” presented at the Proc. Europ. Wireless, Apr. 2007.
%\bibitem{b12} S. Jaeckel, K. Borner, L. Thiele and V. Jungnickel, "A Geometric Polarization Rotation Model for the 3-D Spatial Channel Model," in IEEE Transactions on Antennas and Propagation, vol. 60, no. 12, pp. 5966-5977, Dec. 2012, doi: 10.1109/TAP.2012.2214017.

%%%%theo. Ber
\bibitem{b13} S. Benedetto and P. Poggiolini, "Theory of polarization shift keying modulation," in IEEE Transactions on Communications, vol. 40, no. 4, pp. 708-721, April 1992, doi: 10.1109/26.141426.
%%pathloss
\bibitem{b14} S. W. Ellingson, “Path loss in reconfigurable intelligent surface-enabled channels,” arXiv preprint arXiv:1912.06759, 2019
\end{thebibliography}
\end{document}